# Developing a FPGA – Supported Touchscreen Writing / Drawing System for Educational Environments

**Aslihan Tufekci**
*Gazi University,*
*Gazi Faculty of Education,*
*Dept. of Computer Instruction Technologies, Turkey*
*asli@gazi.edu.tr*

**Kamuran Samanci**
*Turk Telekom, Turkey*
*kamuransamanci@yahoo.com*

**Utku Kose**
*Usak University,*
*Computer Sciences App. & Res. Center, Turkey*
*utku.kose@usak.edu.tr*

## Abstract

Developments in information and communication technologies have been greatly influential on the practices in all fields, and education is not an exception to this. To illustrate with, computers were first used in computer – assisted education in order to increase the efficiency of teaching process. Recently, computer has contributed more to the field through interactive and smart class applications that are specially designed for classroom use. The aim of this study is to develop a low – cost, portable and projection – supported touchscreen to be used in educational environments by using FPGA technology and to test its usability. For the purposes of the study, the above mentioned system was developed by using the necessary hardware and software, and later it was tested in terms of usability. This usability test was administered to teachers, who were the target end – users of this touchscreen writing / drawing system. The aim of this test was to determine "user – friendliness", "subservientness" and "usability" of the system. Several tools were used to obtain data from the users that participated in the study. The analysis and evaluation of the data collected revealed that the system has achieved its objectives successfully.

## 1. INTRODUCTION

Education has always been a field of study making use of technology as effectively as possible. Today's popular concept "educational technologies" has been derived from "the process of benefiting from the advantages provided by technology as much as possible in order to increase the effectiveness and efficiency of teaching process and educational studies. In other words, educational technologies (also called teaching technologies) are defined as "the studies and ethical applications ensuring facilitated education and increased performance by effectively using, managing and developing appropriate technological processes and resources (Richey, 2008).

Technology has always supplemented educational practices in several ways. It is clear that extending and introducing these practices to larger audiences, presenting visual simulations and graphic supports for students, developing various audio – visual educational materials and introducing many evaluation systems and tools result in more efficient and easier educational processes. Similarly, the findings of the studies conducted in educational technologies are to a great extent reflected in real classrooms in the form of practical applications. To illustrate with, many technological tools and devices, most of which are computer – supported, are used for educational processes in real classroom environments. It is also true that almost all the educational materials used in today's classrooms are computer – controlled or computer – operated. Therefore; thanks to computer support, classrooms have now acquired "smart classroom" features. Depending on the developments in computer technology, the users (students or teachers) have started to interact with the devices available in classrooms, which have inevitably led to the emergence of the concept "interactive classroom" as the "next generation" version of "smart classrooms".

In interactive classrooms, there is an ongoing direct interaction between the users and the system, and this process requires a computer enabling this interaction. The studies conducted on this issue show that almost all interactive classroom applications require the use of at least one computer. Thus, the dense use of computers has become inevitable in educational processes just like in every field of life. Used for measurement and evaluation purposes at the beginning, computers later have been used in computer – assisted teaching intensively as well (Dogan, 1997). The most recent example of this new function of computers is computer – supported board, which is also called "smart board". The system is fully controlled by a computer, and the texts and visuals are projected on a board via a projector. These visual elements can be altered by using specially designed pens called "smart magic pen".

The presence of hardware does not suffice for computer – assisted teaching systems. Some necessary software supporting the hardware must also be installed into the system. Developing such software requires hardwork and time, so the total system can be expensive since this hardwork and time spent are taken into consideration while determining the price of the product. Moreover, it is true that computers are multi – purpose devices and not used only for educational purposes. Depending on the particular needs of the customers, various software and hardware – not only for educational purposes – can be installed in standard computers and marketed by computer companies. In other words, the consumers, unfortunately, have to pay for the software and hardware that will not be used in educational processes too. Therefore; the use of computer in a system increases the overall cost considerably. In addition, since computers are complex devices, it is necessary to employ a computer technician at school to deal with the maintenance and repair of the computers.

Based on the explanations mentioned above, the current study aims at the following: Developing simple, low – cost touchscreen writing / drawing system for educational environments which does not require the use of a computer and can be used as portable or stationary; and testing its usability on teachers, who are the real target users of this touchscreen writing / drawing system". In addition, the followings are targeted as secondary objectives:

- To assist teachers, who are the real target end – users, in teaching more easily and give them an alternative to continue their in – class teaching in case of illnesses and even disability since this touchscreen enables them to teach while sitting without needing much physical activity or effort.

- To ensure more visibility of the board – considering the fact that the users might cast a shadow on some parts of the board in the systems like "smart board" depending on the light coming from the projector.

Touchscreen writing / drawing system, which is likely to meet the objectives mentioned above, is expected to be an alternative tool in educational processes and for the systems requiring computers with expensive software and hardware.

## 2. BACKGROUND

Developments and improvements in the context of especially education field have been given rise to appearment of different kinds of educational technologies. As a result of appearment of the related technologies, supportive educational materials for ensuring more quality in teaching – learning processes has also been designed and developed in many scientific research works. At this point, improvements especially within computer and electronic technologies has an important role on rapid developments in the related scope. In this sense; computers, computer – based systems and other technologies from general perspective (especially information and communication technologies) have been widely used in educational studies in performing many developments and improvements (Deperlioglu & Kose, 2013; Kose, 2010; McCormack & Jones, 1997).

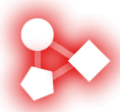

When the current conditions and situations have been examined, it can be seen that usage of computer – based and electronic devices in teaching – learning processes is widely preferred in order to improve quality and effectiveness of the related educational approaches. In this sense, when this issue is evaluated from students' perspective, some research works also report that there is a large range of capacities for especially information and communication technologies among students in higher education (Lee, 2003; Palaigeorgiou *et al.*, 2005; Van Braak, 2004). Related to the discussed issue, such reports also point the situation of remarkable usage popularity of computer and electronic technologies among educational studies.

As being parallel with the research subject of this work, usage of computer and electronic technologies is an important factor that must be taken into consideration to discuss about the background. As it can be understood, such devices based on computer and electronic technologies play active roles on improving teaching – learning processes and ensuring effective and efficient ways for achieving the educational objectives. Related to this issue, it has been argued within some works that the use of such devices (for example, laptops, mobile devices) during teaching – learning processes improves students' grasp and comprehension of the subject by providing flexible interaction and using experiences (Alsaggaf *et al.*, 2012; Dexter *et al.*, 1999; Nilson & Weaver, 2005; Tufekci *et al.*, 2013). On the other hand, it has also been reported that students think very positive about using such materials although some researchers also think that the related devices may affect the learning process in a negative way (Akbaba-Altun, 2006; Barak *et al.*, 2006; Hembrooke & Gay, 2003; Ni & Branch, 2004).

As it was also mentioned before in the "Introduction" section of this article, more advanced computer – assisted materials like "smart board" or "interactive tablet" have also been designed and developed as a result of rapid developments in the mentioned technologies combined in a common scope for improving educational processes. Nowadays, different kinds of educational technology – based materials like smart boards, touchscreen devices, interactive tablets, camera – based viewing devices…etc. are widely used for teaching – learning approaches. Within the current modern life, it can be seen that the touchscreen – based system are used usually as kiosks in order to improve standards in activities in airports, train stations, grocery stores, banks and any other workplaces such as food service, retail and health care fields (Astell *et al.*, 2010; Chourasia *et al.*, 2013; Newman *et al.*, 2012; Schultz *et al.*, 1998; Shervin *et al.*, 2011; Wilson *et al.*, 1995). As being parallel with the expressed points and because of the main objective of this work (designing and developing touchscreen writing / drawing system) and other following objectives; previously performed research studies – works within the subject must also be examined in order to enable readers to have more idea about the work.

In the sense of the current literature, it can be seen that there is not too many studies regarding to design and develop of a touchscreen system for directly educational purposes. Because of this, some recent studies – works employing design, development and usage of touchscreen technologies are also discussed as follows in order to give ideas about touchscreen based applications:

- In their works, Raj *et al.* (2013) provides a research work based on design, development and implementation of a touchscreen health information kiosk. By using this system, patients at St. John's Medical College Hospital – Bangalore have ability to receive information by using kiosk systems provided in the context of specific areas. From the view of receiving information and functions related to especially "feedback" approaches, this system can also be evaluated as some kind of learning – based scientific study providing many benefits in the context of using touchscreen technologies.

- Caviglia-Harris et al. (2012) has performed a study on using computer – assisted data collection approach for providing more accurate and effective way for performing survey – based works. In this study, touchscreen laptop systems have been used for performing surveys and collecting item responses in this way. The study has shown that using the related devices rather than using the traditional method (paper and pencil interviewing) provide a effective way on data collecting by reducing mistakes and ensuring many advantages in the sense of "time and place".

- In his Master Thesis work, Taylor (2012) proposes "a near touch user interface" for touchscreen – based systems. This system has been designed and developed in order to overcome interaction limitations of traditional touchscreen applications. In this sense, Taylor uses some supportive devices to form the proposed touchscreen system – approach.

- In the context of medical applications, Aguirre *et al.* (2012) provides a system developed as a touchscreen – based, low cost electrocardiograph. The results obtained with the work show that the system also provides a well performance within applications of the electrocardiogram.

- Dixon and Prior (2012) have provided a research work on a typical touchscreen electronic queuing system. In this scope, they have provided a work on "design and initial usability testing of the anonymous electronic waiting room system" for improving patient satisfaction. The obtained results show that the system has positive affects on patients and provided successful rates on other related factors like "interaction".

- Related to usage of touchscreen – based systems, Chourasia *et al.* (2013) have provided a work on "evaluating effects of sitting and standing on performance and touch characteristics during a digit entry touchscreen task in individuals with and without motor – control disabilities. The work provides a good evaluation approach for evaluating different factors (for example button sizes of a touchscreen) in order discuss about the related effects and generally the results points that "environmental conditions should also be considered to improve accessibility and usability of touchscreen". As being similar to this work, some previous works on effect of touch screen interface design on performance have also been provided (Colle & Hiszem, 2004; Jin *et al.* 2007; Martin, 1988; Sesto *et al.* 2012)

- In their work, Rahman *et al.* (2012) provides a touchscreen – based automation system to control electronic devices within a home. In the context of the proposed system, electronic components that can be bought from any available local market have been used in order to form a low cost device.

- Another remarkable work regarding to development of touchscreen approach has been provided by Bi *et al.* (2012). In this work, a multilingual touchscreen – based keyboard application has been designed and developed by the authors. In this sense, the keyboard button layout has been optimized according to usage of the keyboard for different languages like French, Spanish, German, and Chinese.

- Regarding to optimization of touchscreen systems and applications, another work has been performed by Bradley (2012), in order to determine "factors affecting the adoption of touchscreen smartphones among individiuals with vision loss". According to the obtained results within this Master Thesis work, more user – friendly, assistive technologies can be performed on mobile devices, in order to fit them to be used better by especially individiuals with vision loss.

After the taking a brief look at to the background and the literature status, the touchscreen writing / drawing system, which was provided within this work must be examined in the context of design process. In this way, the low cost aspects of the system and its using features and functions can also be understood better.

## 3. DESIGN PROCESS OF TOUCHSCREEN WRITING / DRAWING SYSTEM

As it was also mentioned before, a low cost projection – supported touchscreen writing / drawing system, which does not require a computer and can be used as portable or stationary, was developed in the current study by using the advantages provided by FPGA technology. In this sense, Figure 1 represents some fotographs taken from the designed and developed system (as prototype).

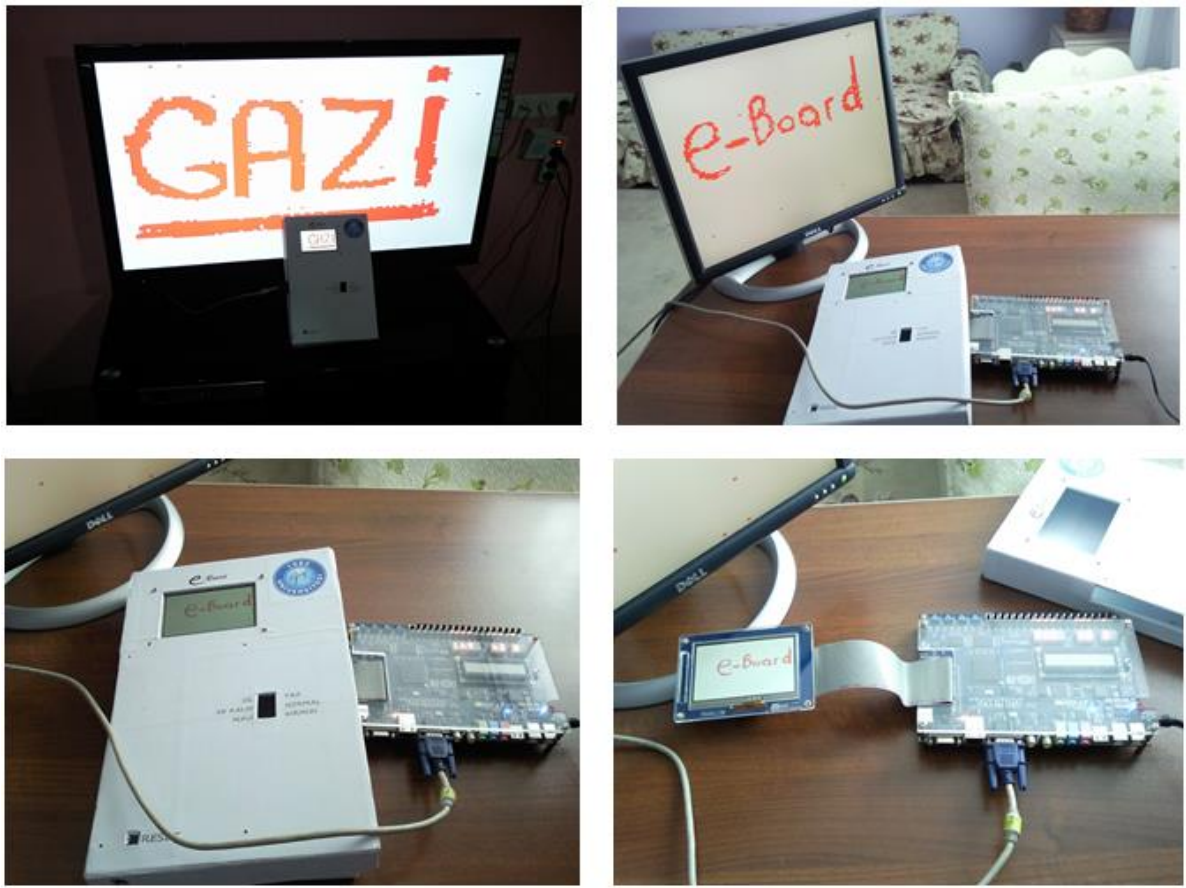

**Figure 1.** Some fotographs taken from the designed and developed system (as prototype).

Regarding to the touchscreen writing / drawing system, it is necessary to define the design and development process of "touchscreen writing / drawing system" in terms of software and hardware in order to learn more about the study and have a clearer idea of its importance. By doing so, the features and functions of the system are also explained in detail. In order not to confuse the readers, technical details were omitted from the explanations.

### 3.1. Hardware Design of Touchscreen Writing / Drawing System

The working mechanism of the system hardware is reprensented in the diagram under Figure 2. This figure also presents the preview of the system software.

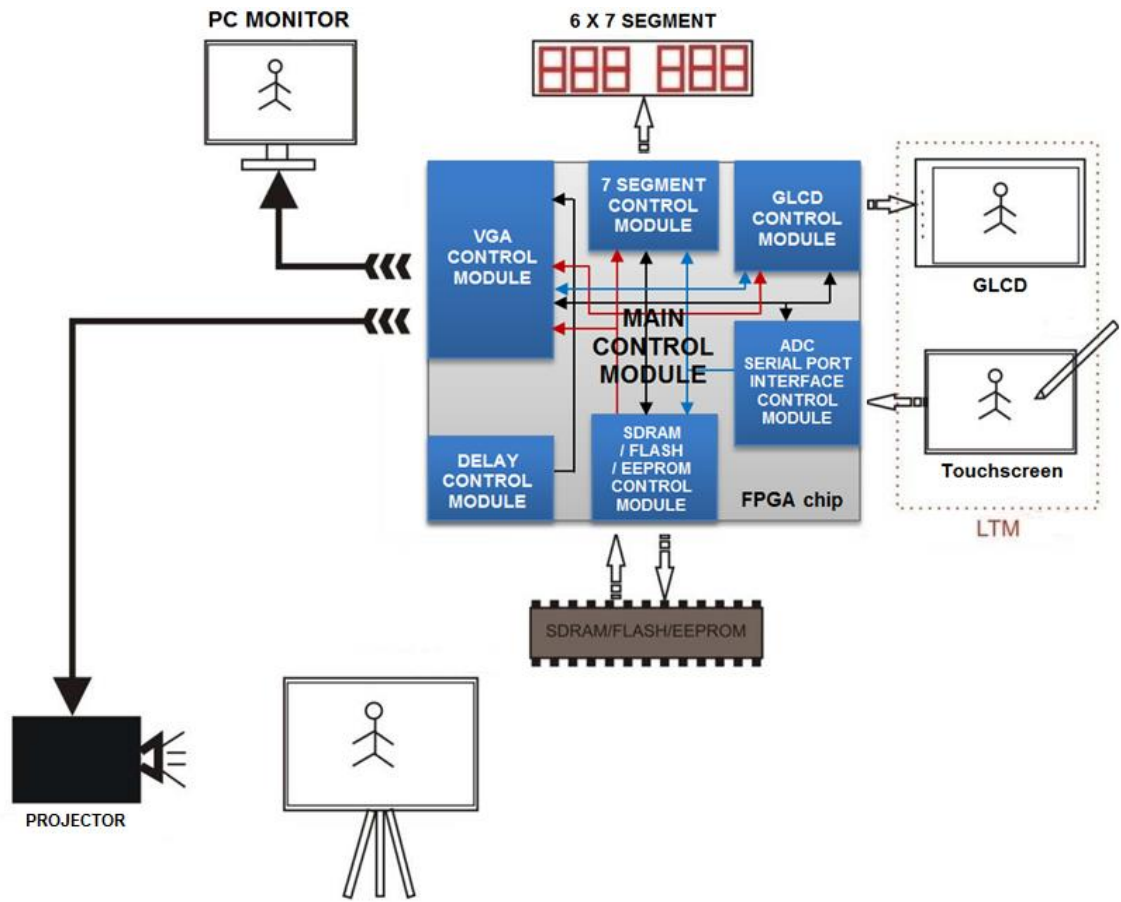

**Figure 2.** System structure of the touchscreen writing / drawing system.

As seen in Figure 3.2, X and Y coordinates of the writings or drawings made on the panel with the help of a special pen and a touch – operated panel located on touchscreen GLCD panel (LTM – LCD Touch Module) are converted into digital values through an ADC (Analog to Digital Converter) chip. These digitally converted X and Y coordinate values are transferred to FPGA content through "ADC serial port interface (SPI) control module" which is located on FPGA chip. These transferred values are stored in an internal register RAM (Random Access Memory) via "SDRAM or Register control module" located on FPGA chip. Internal register is used in this current system. Coordinate information stored in the register again via the same module, are synchronously sent to "GLCD control module", "7 – segment display module" and "VGA control module". "GLCD control module" is integrated into FPGA chip in order to control touchscreen GLCD panel. This module processes the coordinate information receieved from "SDRAM or Register Control Module" and later enables this processed information to be converted into images. "7 – segment display control module" is used to control six "7 – segment diplays". This module seperates X and Y coordinate values as 3 – segment displays that are recieved from "SDRAM or Register Control Module" and realizes the visual presentation. On the other hand, "VGA Control Module" is used to transfer the images to a computer screen (or a projector device) to be connected to main board with the help of a VGA (Video Graphics Array) chip. In short, the function of VGA control module is to process the coordinate information receieved from "SDRAM or Register Control Module" and to transfer the processes information to computer screen.

 "Touchscreen Writing / Drawing System" is composed of two hardware units, one of which is "LCD touchscreen panel" module (also called LTM) and the other is "control" module formed with FPGA chip.

### 3.1.1. LCD touchscreen panel module

As seen in Figure 3, LCD Touch Panel Module (LTM) includes a 4.3 inch graphic LCD with 800X400 pixels and 24 – bit color resolution and also 4.3 inch resistive touchscreen.

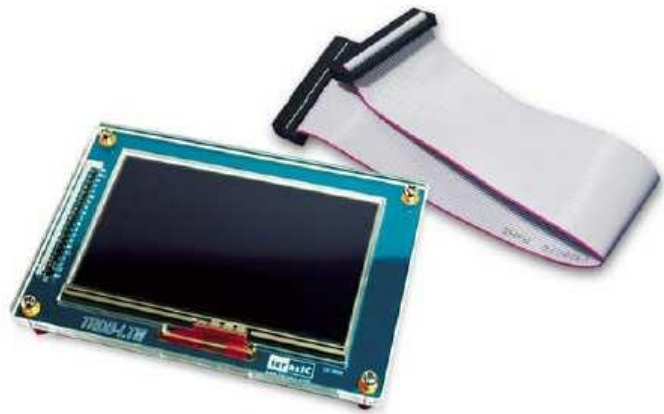

**Figure 3** LTM module and the cable used for connection to main board.

This module is mainly used for system control and image transfer function. However, it can be used for various applications as well with its high resolution screen and touch panel support. The aim and the function of LTM module in this touchscreen writing / drawing system are as follows: Firstly, the things written or drawn by a special pen on the "active area" of the module are converted in to 12 – bit X and Y coordinate values via an ADC chip to which touch panel is connected. Figure 4 shows the "active area" of the module and 12 – bit X and Y coordinates.

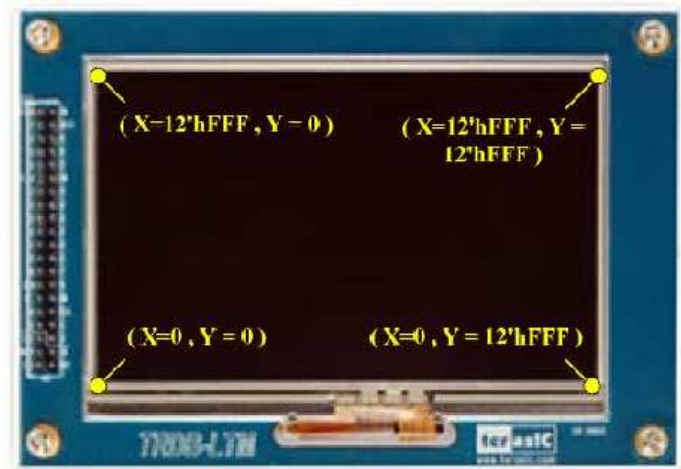

**Figure 4.** The display of coordinates X and Y on LTM module as 12 – byte groups.

The connection between LTM panel and the main system is made by 40 – pin socket connection. The connection of digitally converted X and Y coordinate values to the main board, which also includes the FPGA chip, is also made via this socket (Figure 5).

Secondly, the data converted into meaningful formats in FPGA chip is transferred back to the GLCD screen found on that module. By doing so, the visual transfer of the drawings made on touch panel is realized (Figure 5).

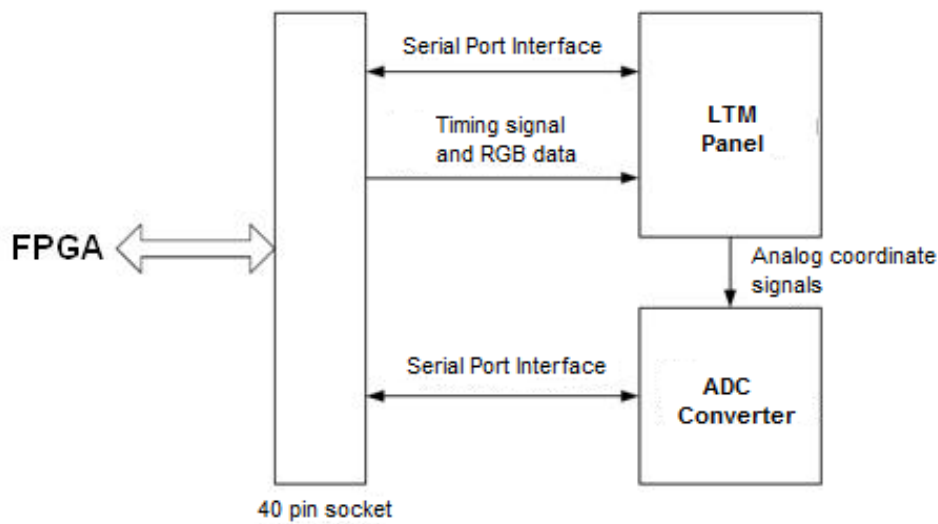

**Figure 5.** Data flow diagram on LTM panel.

### 3.1.2. Control module

The control module deals with the following: FPGA chip, which receives X and Y coordinates from GLCD touch panel and later enables the conversion of these processed vaules into images on the LCD display of GLCD touch panel as well as on a computer screen or projection device simultaneously; and hardware connections of other chips which are involved in the process as supporting components. Figure 6 shows FPGA chip – which is the main component of touchscreen system main board – and other supplementary chips, sockets and connections.

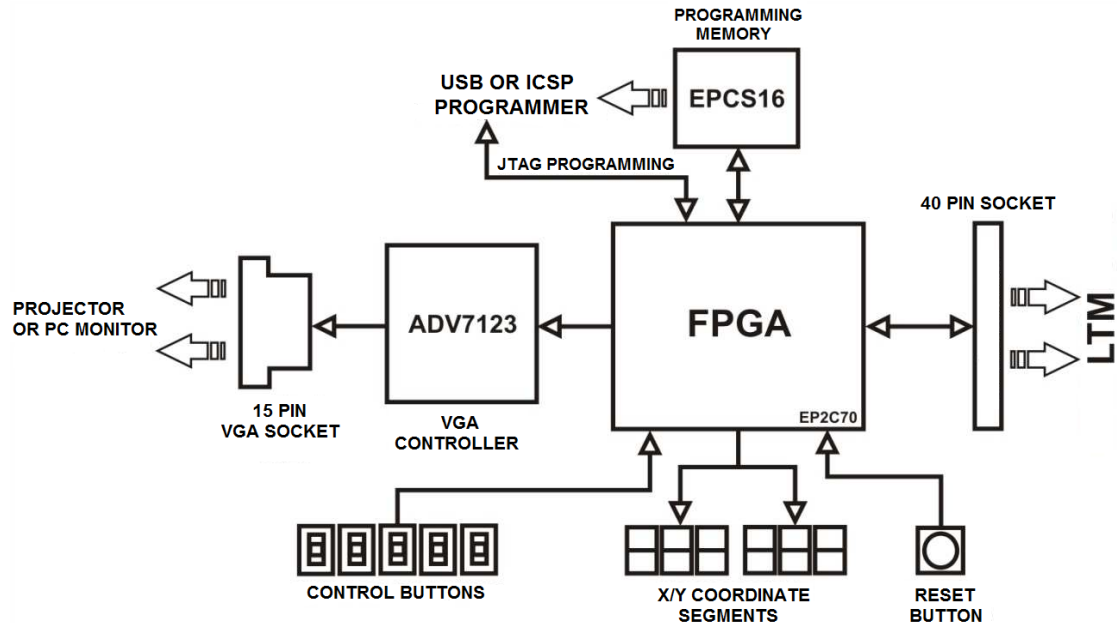

**Figure 6.** Connections between the chip, sockets and other components on the system main board.

The FPGA chip used in touchscreen writing / drawing system developed for the purposes of this current study is called "EP2C70896C6N chip", which is the most advanced chip of Cyclone II family produced by Altera company.

- It is produced by using 90 nm technology.

- It includes 68,416 logical elements (LEs).

- Mbit embedded RAM.

- Performance upto 260 MHz.

- Compatible with high – speed mobile external disks such as DDR, DDR2 and SDR SDRAM.

- Multi – volt multiple input/output voltage support (1.5V, 1.8V, 2.5V, 3.3V).

- Configuration in less than 100ms thanks to quick – configuration opition.

- 16 sensitive clock input pins.

- A total of 150 "18 x 18 multiplexer".

- It supports configurations in Active serial, Passive serial and JTAG mode.

▪ 896 – pin BGA socket structured.

"EP2C70896C6N" chip is found to be the ideal one when the features mentioned above and the purposes of the study are considered. At this point, learning about software as well hardware design of the touchscreen writing / drawing system is also important in terms of being familiarized with its features and the operation process.

## 3.2. Software Design of Touchscreen Writing / Drawing System

Software design of touchscreen writing / drawing system mainly depends on the programming of FPGA chip. It is clear that effective and efficient operation of the system according to the predetermined purposes will be possible only if the chip is programmed appropriately, that is software infrastructure should be designed and integrated into the chip. This process also refers to the configuration of FPGA chip.

As for the programming of FPGA chip, Verilog software programming language was used. In addition, Quartus II software, produced by Altera Company, was used for the same purpose. In terms of software, FPGA chip is composed of a total of 7 software modules (Figure 7).

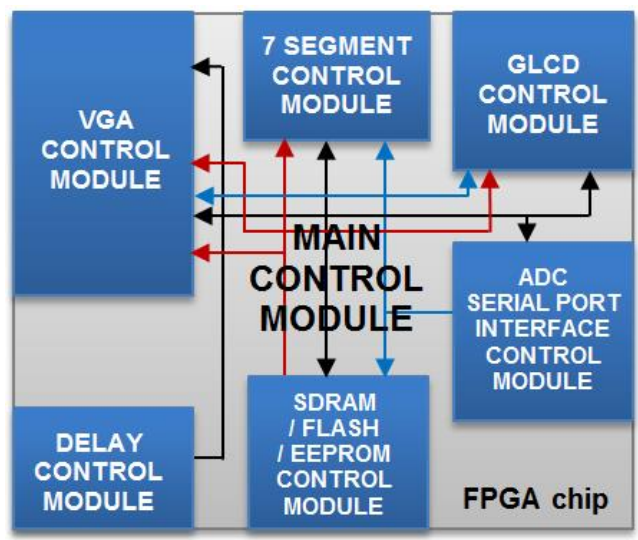

**Figure 7.** Software modules on FPGA chip.

The basic features and functions of software modules can be explained briefly as follows:

### 3.2.1. Main control module

Main control module refers to the main parts of software system on FPGA chip. It covers all software modules in the chip and enables the connections among the modules. In addition, it is responsible for the connections of the modules with external world and the necessary controls accordingly.

### 3.2.2. Delay control module

Delay control module enables the modules on the main module to be processed after a certain delay time. The aim of this function is to ensure harmonius operation between software and hardware processes and to avoid a potential system error accordingly.

### 3.2.3. ADC serial port interface control module

ADC serial port interface control module is responsible for the software controls related to ADC serial port interface communication during the system operation process with regards to FPGA chip.

### 3.2.4. Graphic LCD control module

GLCD control module is the software module enabling data transfer between LCG driver chip and FPGA chip by establishing a protocol necessary for this transfer. This module consists of two sub – modules. One of them, which is "GLCD SPI Control Sub – module", was formed to create a common language between LCD driver chip and FPGA by forming the necessary protocol for the communication between these two chips. The other sub – module – GLCD Timing sub – module, is a software module that determines the quality, size, resolution and speed of the data to be sent to GLCD.

### 3.2.5. VGA control module

VGA control module is a software module enabling VGA signal synchronization on both vertical and horizontal coordinates in order to obtain the desired image levels in the devices such as projector or monitor that are connected to the system via VGA.

### 3.2.6. Seven – segment control module

This module is the software module that realizes the processes necessary to carry out accurate transfer of related data and to display it on a "seven – segment display group". Here the aim is to convert the coordinate data of the points touched on touch panel on X and Y coordinates.

### 3.2.7. SDRAM / Flash / EEPROM control module

Being the last software module in FPGA, SDRAM / Flash / EEPROM control module is responsible for the communication between memory units and the system depending on which "touchscreen writing / drawing system" is used.

Touchscreen writing / drawing system, whose software and hardware design has been explained without getting into technical details, was planned with an approach to enable the users to complete the actions effectively based on the purposes of the study. At this point, it is necessary to make an evaluation to determine the effectiveness of the system and its adequacy to meet the purposes of the study.  To achieve this aim, the following evaluation process was planned.

## 4. EVALUATION PROCESS

The evaluation process of touchscreen writing / drawing system mentioned in this study included the following phases; observing usability processes and the analysis of the obtained opinion data. Accordingly, the details related to evaluation process will be presented in the following sub – titles and paragraphs.

### 4.1. Evaluation Method

In order to evaluate touchscreen writing / drawing system, an approach focusing on the usability of the system was preferred. In order to achieve this purpose, Heuristic Evaluation Method was used. This method is generally applied for the usability evaluation of the systems that are based on learning through trial – error (Nielsen & Molich, 1990).

In this study – specific touchscreen writing / drawing system, there are not any written documents except the names of the buttons on the device activating the related functions. In other words, the information about the system usage is limited to only this text – based information. According to Nielsen, a total of 5 subjects are sufficient to test such systems since 75 % of the problems can be determined by that number of participants (Nielsen & Landauer, 1993). Similarly, in the related studies conducted in the field, the number of the problems identified was not found to increase as the number of the participants increased. The graph in Figure 8 represents the problem identification data according to the number of participants in the studies conducted by Nielsen and Landauer between 1990 and 1993 (Nielsen & Landauer, 1993).

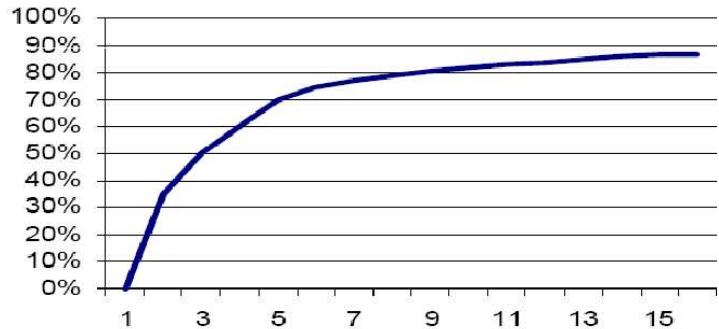

**Figure 8.** Problem identification rate according to the number of the subjects (Nielsen, & Landauer, 1993).

However; the study conducted by Faulkner (2003) shows that five participants might not suffice. In the study, it was found that the problems identified by the sub groups (five participants) formed out of a larger group of 60 participants corresponds to 85% of the problems identified by the whole group. As the test simulations applied increased, the usability problems identified by groups of five were found to have a very wide range of match or mismatch (between 55 % and 100 %) with the problems identified by the whole group. This wide deviation implies that the potential problems are very likely to be missed when groups of five participants are used in similar usability studies. However, using groups of ten participants increases this percentage to the average 95 %, and 82 % being the lowest value. Table 1 shows the changes in standard deviation in brief.

**Table 1.** The changes in standard deviation with regards to the related variables.

| User Number | Determined Minimum % | Determined Mean % | Std. Deviation |
|:---:|:---:|:---:|:---:|
| 5 | 55 | 85,55 | 9,30 |
| 10 | 82 | 94,69 | 3,22 |
| 15 | 90 | 97,05 | 2,12 |
| 20 | 95 | 98,40 | 1,61 |
| 30 | 97 | 99,00 | 1,13 |
| 40 | 98 | 99,60 | 0,81 |
| 50 | 98 | 100,00 | 0,00 |

When above mentioned data is concerned, a group of 10 users as the participants of the study can be considered to be better in identifying the usability problems.

## 4.2. The Population and Sample for Evaluation

The population of the current study involves teachers working in primary schools regardless of their fields of teaching. On the other hand, the sample of this study is "randomly selected 20 teachers" from various fields of teaching who taught during 2011 – 2012 academic year.

## 4.3. Evaluation Material

Due to the limited number of similar studies in the related literature, the authors developed a survey in order to test the usability of this touchscreen writing/drawing system. After several prior analyses, a 35 – item list was prepared by the authors. Later, these items were reviewed and edited by an expert and a 15 – item evaluation survey form was finalized, which aims at determining "subservientness", "user – friendliness" and "usability" factors regarding the system. The distribution of these 15 items according to the factors is as follows: first five 5 items for subservientness, the next 5 for user – friendliness and the last 5 for usability. The participants of the study are asked to choose one of three options for each item; namely "I agree (Yes)", "I partly agree (Partly)" and "I don't agree (No)".

## 4.4. Other Approaches used in the Evaluation

Prior to the survey used as part of the evaluation approach, the participants were asked to fill out an information form in order to learn about their familiarity with the study topic. The data obtained from this phase revealed that the participants had enough knowledge about the topic; and therefore, the replies to the survey items were considered to be quality enough for the purposes of the study. The details of this form will not be provided here so as not to confuse the readers and to focus on the survey study more.

Another approach used for the evaluation plan is the observation of usability processes, which is based on completing the tasks given regarding the use of touchscreen writing / drawing system designed and developed for the purposes of this study. Although the survey study seemed to have the primary importance with regards to evaluation, this phase was also crucial since it provided valuable data about the use of the system before the participants filled out the survey. The next section presents the data obtained regarding the approach mentioned above.

## 5. FINDINGS OF THE EVALUATION PROCESS

As mentioned in the previous section, the process was observed and the participants were administered the survey developed by the researchers, and later the necessary data was collected for the evaluation purposes of touchscreen writing / drawing system designed and developed for the purposes of the current study. Therefore; the data obtained will be presented in the next sections in the order followed while collecting the data; namely the data regarding the observations and the data obtained from the survey study respectively.

### 5.1. Findings Obtained from Usability Process

The usability process deals with the time the participants of the study spent while completing the tasks given as well as the problems they encountered during this completion process. The participants were not provided any information about the use of touchscreen writing / drawing system prior to the process. They were only given information about the objectives of the system and brief introduction to its working principles.

As part of the process, the participants were given a list including a total of 8 tasks. The time they spent while working on the tasks was recorded by an observer as seconds. In addition, each participant was asked to decide on the difficulty level of each task by marking a "five point Likert scale" available below each task in the list.

The information and the findings obtained from this process are presented below:

As shown in Table 2, the participants spent an average of 19 seconds for each task and the task list, which includes 8 tasks, was completed in an average of 3 minutes by per participant. The task that was completed in the shortest time (15.1 second) was "clear the screen and shut down the sytem" (Task 8), while the longest time spent was 23,60 seconds for the task which writes: "connect touch panel to the screen and switch it on in adaptor mode" (Task 1). The reason why Task 1 had the longest completion time might be explained by the fact that the participants used such a system for the first time. In other words, it might have been due to effect of first time experience and momentary panic and nervousness.

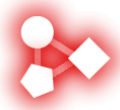

**Table 2.** Usability process – task completion times (P1, P2….. P20= participants).

| Task | Task Completion Times (sec.) | | | | | | | | | | | | | | | | | | | | Mean |
|---|---|---|---|---|---|---|---|---|---|---|---|---|---|---|---|---|---|---|---|---|---|
| | P1 | P2 | P3 | P4 | P5 | P6 | P7 | P8 | P9 | P10 | P11 | P12 | P13 | P14 | P15 | P16 | P17 | P18 | P19 | P20 | |
| 1. Connect touch panel to the screen and switch it on in "Adaptor" mode. | 23 | 24 | 23 | **41** | 13 | 27 | 34 | 16 | 14 | **10** | 19 | 22 | 27 | 25 | 30 | 23 | 18 | 23 | 31 | 29 | **23,60** |
| 2. After touch screen is displayed, shut it down and restart in "Battery" mode. | 27 | 34 | 35 | **38** | 19 | 23 | 27 | 21 | 21 | **19** | 21 | 18 | 16 | 21 | 17 | 20 | 13 | 18 | 15 | 16 | 21,95 |
| 3. Do some drawings on the screen when "draw" mode is on. | 10 | 19 | 15 | **25** | 17 | 12 | 15 | 18 | 7 | **12** | 17 | 26 | 21 | 14 | 23 | 16 | 19 | 17 | 20 | 18 | 17,05 |
| 4. Switch to "Erase" mode and erase what you have drawn. | 15 | 21 | 19 | **23** | 17 | 18 | 16 | 9 | 12 | **10** | 14 | 19 | 23 | 32 | 28 | 17 | 16 | 26 | 18 | 22 | 18,75 |
| 5. Do drawings on the screen when "Draw Bold" mode is on. | 18 | 18 | 26 | **32** | 23 | 16 | 19 | 16 | 15 | **13** | 16 | 21 | 18 | 17 | 22 | 19 | 15 | 22 | 15 | 17 | 18,90 |
| 6. switch to "Erase Bold" mode and erase what you have drawn. | 11 | 27 | 24 | **15** | 19 | 16 | 13 | 11 | 11 | **11** | 20 | 18 | 15 | 21 | 17 | 22 | 19 | 27 | 23 | 26 | 18,30 |
| 7. Change the font color from "Red" to "Blue". | 17 | 15 | 20 | **21** | 21 | 17 | 14 | 8 | 14 | **10** | 23 | 21 | 19 | 25 | 16 | 18 | 22 | 14 | 17 | 15 | 17,35 |
| 8. Clear the screen and shut down the system. | 15 | 14 | 12 | **19** | 14 | 12 | 9 | 7 | 13 | **12** | 19 | 20 | 22 | 15 | 17 | 13 | 18 | 19 | 18 | 14 | **15,10** |
| **Total:** | 136 | 172 | 174 | **214** | 143 | 141 | 147 | 106 | 107 | **97** | 149 | 165 | 161 | 100 | 108 | 140 | 146 | 167 | 157 | | |
| **Overall Mean:** | | | | | | | | | | | | | | | | | | | | | 18,87 |

The Figure 9 represents the graph of the average times spent on task completion as part of evaluation process.

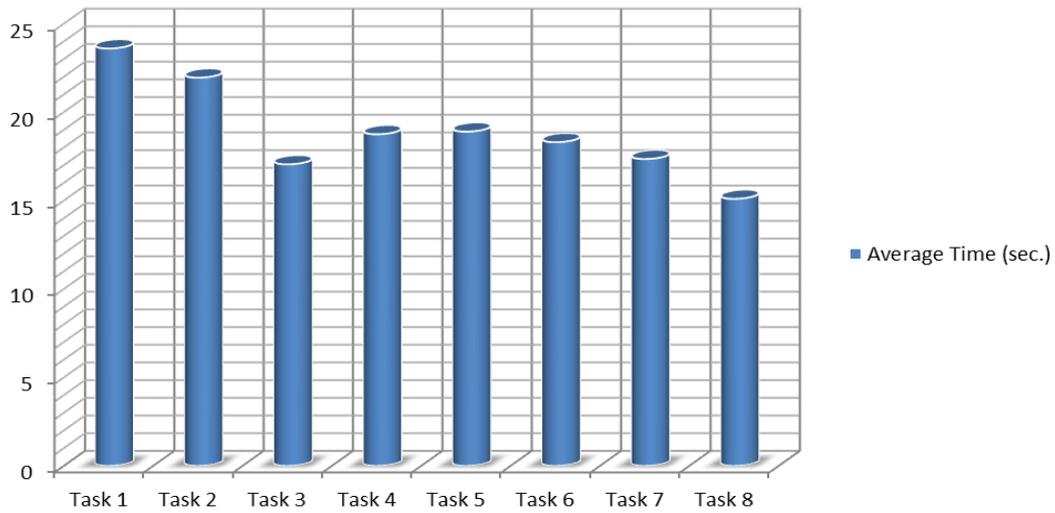

**Figure 9.** Usability process – average task completion times.

Table 3 shows the data regarding the difficulty level of the tasks as stated by the participants during the process. When the table is examined, it is seen that the mean value for all tasks is 3,75 (easy). The most difficult task was found to be Task 2 "After touch panel is switched on, switch it off and restart in battery mode" with an average of 3,15 while the easiest task was Task 4 with a balue of 4,45 which writes "clear all drawings from touch panel" by using the "clear mode".

**Table 3.** Usability process – difficulty levels of the tasks (P1, P2..... P20= participants).

| Task | Difficulty levels of the tasks (1:very difficult; 2:difficult; 3:almost easy; 4:easy, 5:very easy) | | | | | | | | | | | | | | | | | | | | Mean |
|---|---|---|---|---|---|---|---|---|---|---|---|---|---|---|---|---|---|---|---|---|---|
| | P1 | P2 | P3 | P4 | P5 | P6 | P7 | P8 | P9 | P10 | P11 | P12 | P13 | P14 | P15 | P16 | P17 | P18 | P19 | P20 | |
| 1. Connect touch panel to the screen and switch it on in "Adaptor" mode. | 3 | 4 | 3 | 4 | 3 | 4 | 2 | 3 | 2 | 4 | 4 | 3 | 4 | 3 | 4 | 2 | 3 | 2 | 4 | 4 | 3,25 |
| 2. After touch screen is displayed, shut it down and restart in "Battery" mode. | 2 | 3 | 3 | 4 | 3 | 4 | 4 | 3 | 3 | 4 | 2 | 3 | 4 | 3 | 2 | 4 | 3 | 2 | 4 | 3 | **3,15** |
| 3. Do some drawings on the screen when "Draw" mode is on. | 5 | 4 | 5 | 4 | 4 | 4 | 4 | 3 | 3 | 5 | 4 | 5 | 4 | 4 | 4 | 3 | 3 | 5 | 4 | | 4,05 |
| 4. Switch to "Erase" mode and erase what you have drawn | 5 | 5 | 5 | 4 | 5 | 4 | 5 | 4 | 3 | 5 | 5 | 5 | 4 | 5 | 4 | 5 | 4 | 3 | 5 | 4 | **4,45** |
| 5. Do drawings on the screen when "Draw Bold" mode is on. | 3 | 3 | 4 | 3 | 4 | 3 | 4 | 4 | 4 | 5 | 3 | 4 | 3 | 4 | 3 | 4 | 4 | 4 | 5 | 3 | 3,70 |
| 6. Switch to "Erase Bold" mode and erase what you have drawn. | 3 | 4 | 5 | 4 | 4 | 4 | 5 | 4 | 5 | 5 | 4 | 5 | 4 | 4 | 4 | 5 | 4 | 5 | 5 | 3 | 4,30 |
| 7. Change the font color from "Red" to "Blue". | 4 | 4 | 4 | 4 | 4 | 3 | 4 | 3 | 4 | 5 | 4 | 3 | 4 | 3 | 4 | 3 | 4 | 3 | 4 | 5 | 3,80 |
| 8. Clear the screen and shut down the system. | 3 | 4 | 3 | 4 | 3 | 3 | 4 | 4 | 3 | 3 | 4 | 3 | 4 | 3 | 3 | 4 | 4 | 3 | 3 | 4 | 3,45 |
| **Total:** | 28 | 31 | 32 | 31 | 30 | 29 | 32 | 28 | 27 | 36 | 30 | 31 | 31 | 30 | 27 | 32 | 28 | 26 | 36 | 28 | |
| **Overall Mean:** | | | | | | | | | | | | | | | | | | | | | 3,76 |

For all the tasks in the process, the difficulty levels stated by the participants are showed in Figure 10, which shows that Task 4 was the easiest one and Task 2 the most difficult.

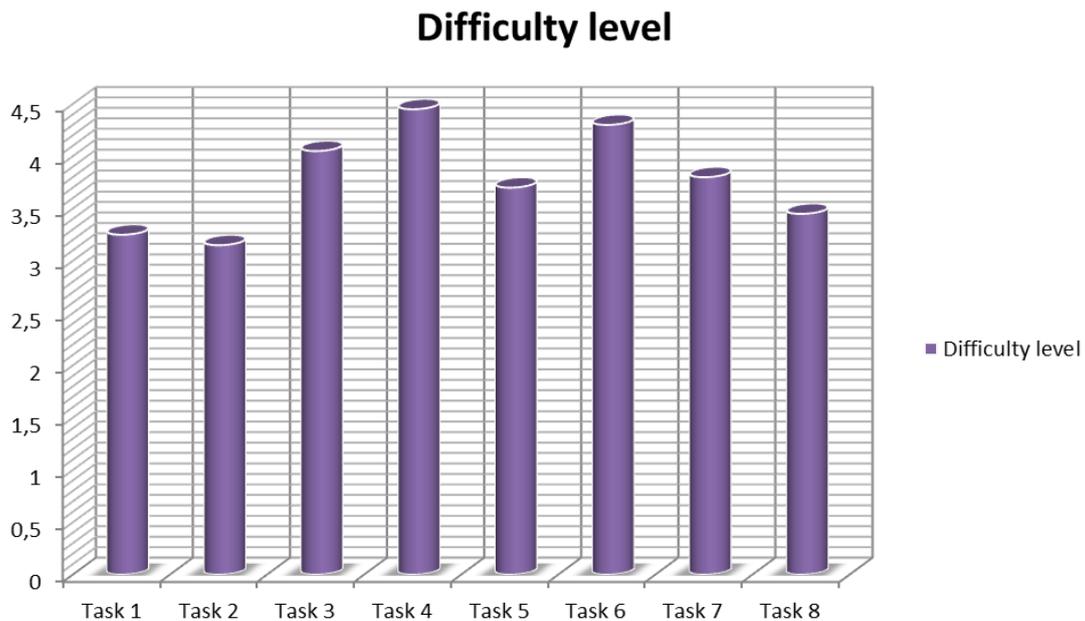

**Figure 10.** Usability process – difficulty levels of the tasks (1:very difficult; 2: difficult; 3:almost easy; 4: easy, 5: very easy).

### 5.2 Usability –Findings Obtained from the Survey Study

As mentioned earlier, a total of 20 participants were administered a 15 – item survey after the usability process, which aims at evaluating the touchscreen writing / drawing system in terms of subvertiness, user – friendliness and usability. In this regard, basic information about the approach and the findings obtained are presented as follows:

As mentioned earlier, the reply options for the items in the survey were "Yes", "Partly" and "No". The answers given to the questions in the survey were analyzed through frequency (f), percentage (%), and standard deviation calculations. The items and the replies – feedback given to these items are presented in the tables in the following pharagraphs.

**Table 4.** The analysis of the replies provided for the statements determining the "subservientness" level of the touchscreen writing / drawing system.

| Statements<br>Do you think that… | No (1) | | Partly (2) | | Yes (3) | | Total | | Mean |
|---|---|---|---|---|---|---|---|---|---|
| | f | % | f | % | f | % | f | % | |
| …you can teach your lessons more easily and quickly by using the "touchscreen writing/drawing system"? | 2 | 10 | 7 | 35 | 11 | 55 | 20 | 100 | 2,45 |
| …your current teaching performance might increase by using the "touchscreen writing/drawing system"? | 2 | 10 | 5 | 25 | 13 | 65 | 20 | 100 | 2,55 |
| …your teaching efficiency might increase by using the "touchscreen writing/drawing system"? | 3 | 15 | 5 | 25 | 12 | 60 | 20 | 100 | 2,45 |
| …you will be more helpful to your students by using the ""touchscreen writing/drawing system"? | 4 | 20 | 6 | 30 | 10 | 50 | 20 | 100 | 2,30 |
| …the "touchscreen writing/drawing system"is a useful tool for your lessons? | 1 | 5 | 8 | 40 | 11 | 55 | 20 | 100 | 2,50 |
| **Overall Mean:** | 2,4 | 12 | 6,2 | 31 | 11,4 | 57 | 20 | 100 | 2,45 |
| No (1) 1,00 – 1,66 %0 – %33<br>Partly (2) 1,67 – 2,33 %34 – %66<br>Yes (3) 2,34 – 3,00 %67 – %100 | | | | | | | | | |

When Table 4 is examined and overall mean values are considered, it is seen that subservientness was given a high (subservient) value by the participants; that is 11,4 the highest out of 20 (57 %). Similarly, the overall mean (2,45) shows that the system is subservient since the value is between 2,34 and 3,00.

When Table 4 is examined in terms of the items, the following implications can be made: teachers can teach their lessons more easily and quickly; their performances and their teaching efficiency can be affected; touchscreen writing / drawing system will be beneficial for their students; and this system will be a useful teaching tool.

The graphs showing the data regarding subservientness evaluation results are presented under the Figure 11.

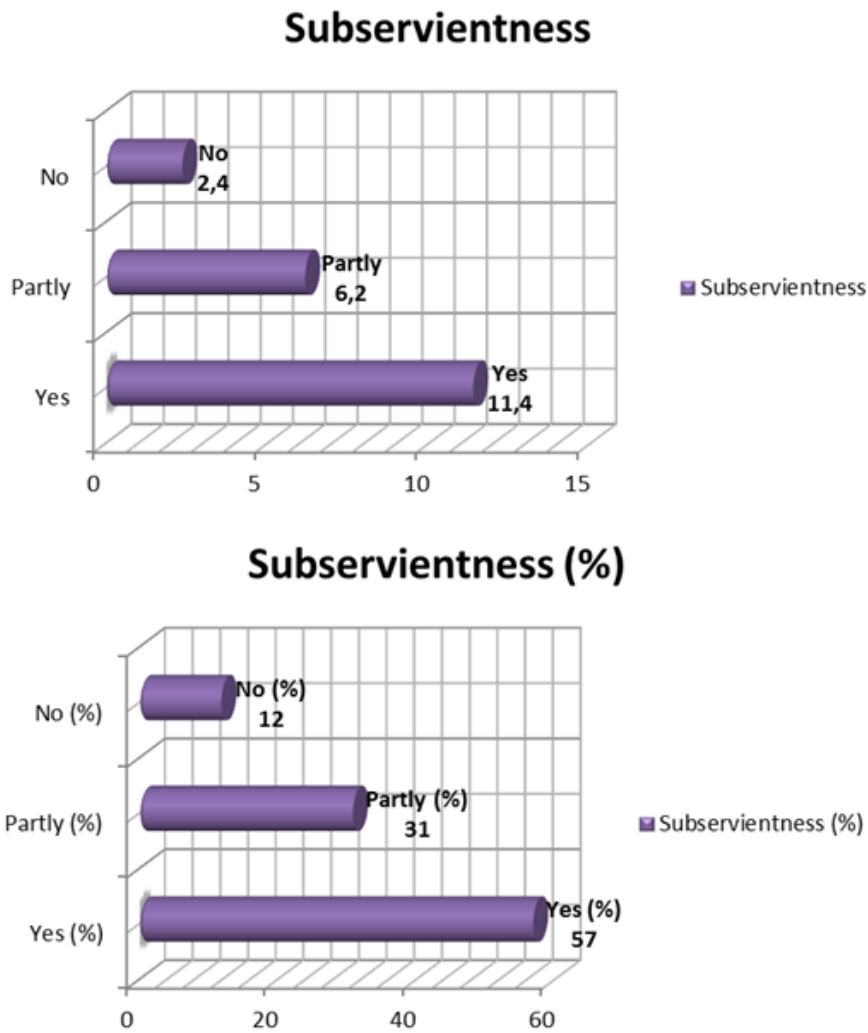

**Figure 11.** Usability survey study – average values with regards to "subservientness" evaluation results.

The results of the survey with regards to "user – friendliness" factor are represented in Table 5. When Table 5 is examined and overall mean values are considered, the "user – friendliness" of the system is found to be evaluated as "easy" with the highest value 11,2 out of 20 (56%). Similarly, the overall mean with a value of 2,37 is between 2,34 and 3 points, which shows that "user – friendliness" of the system is "Yes, easy".

When Table 5 is examined regarding to the items, the following can be concluded: touch panel is user – friendly and easy – to – learn; it is easy to teach something by using the "touchscreen writing/drawing system"; interaction with the screen is clear and simple; and finally it is easy to do drawings on the screen.

**Table 5.** The analysis of the replies to the statements aiming at determining the "user – friendliness" level of the "touchscreen writing / drawing system".

| Statements | No (1) | | Partly (2) | | Yes (3) | | Total | | Mean |
|---|---|---|---|---|---|---|---|---|---|
| | f | % | f | % | f | % | f | % | |
| Did you find it easy to use the "touchscreen writing/drawing system"? | 3 | 15 | 4 | 20 | 13 | 65 | 20 | 100 | 2,50 |
| Was it easy to learn how "touchscreen writing/drawing system" worked? | 4 | 20 | 5 | 25 | 11 | 55 | 20 | 100 | 2,35 |
| Do you think it will be easy to teach something by using "touchscreen writing/drawing system"? | 5 | 25 | 5 | 25 | 10 | 50 | 20 | 100 | 2,25 |
| Do you think the interaction with "touchscreen writing/drawing system" is clear and simple (not complex)? | 2 | 10 | 5 | 25 | 13 | 65 | 20 | 100 | 2,55 |
| Was it easy to make drawings you wanted to on "touchscreen writing/drawing system"? | 5 | 25 | 6 | 30 | 9 | 45 | 20 | 100 | 2,20 |
| **Overall Mean:** | 3,8 | 19 | 5 | 25 | 11,2 | 56 | 20 | 100 | 2,37 |
| No (1) 1,00 – 1,66 %0 – %33 Partly (2) 1,67 – 2,33 %34 – %66 Yes (3) 2,34 – 3,00 %67 – %100 | | | | | | | | | |

The graphs showing mean values regarding "user – friendliness" are presented under Figure 12.

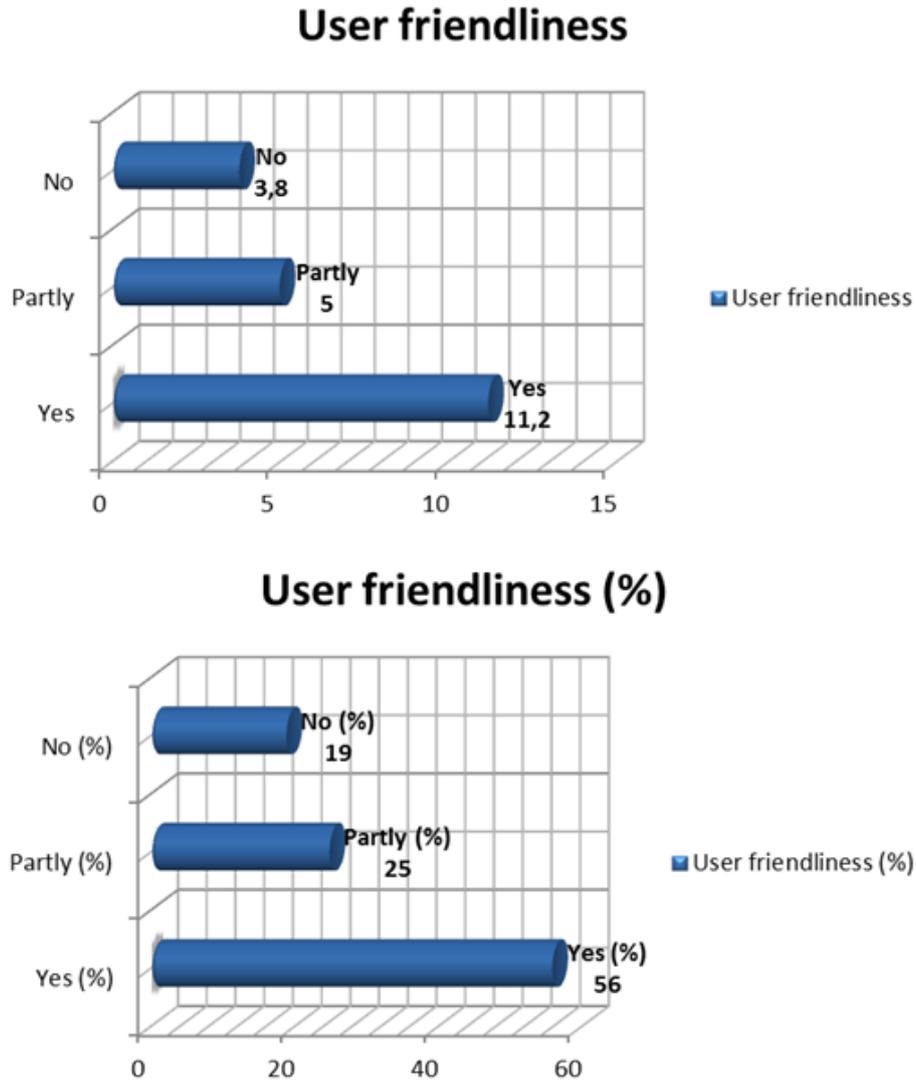

**Figure 12.** Usability survey study – The average values regarding the findings of "user – friendliness" evaluation.

The results of the survey with regards to "usability" factor are showed in Table 6. When Table 6 is examined and overall mean values are considered, the "usability" of the system is found to be evaluated as "partly usable" with the highest value 10 out of 20 (50%). Similarly, the overall mean with a value of 2,36 is between 2,34 and 3 points, which show that the usability level of "touchscreen writing/drawing system" is "Yes, it is usable".

**Table 6.** The analysis of the replies to the statements aiming at determining the "usability" levels of touchscreen writing / drawing system.

| Statements | No (1) | | Partly (2) | | Yes (3) | | Total | | Mean |
|---|---|---|---|---|---|---|---|---|---|
| | f | % | f | % | f | % | f | % | |
| Do you think"touchscreen writing/drawing system" is useful for teaching your lesson? | 2 | 10 | 6 | 30 | 12 | 60 | 20 | 100 | 2,50 |
| Do you think many features of the product are necessary and appropriate? | 1 | 5 | 8 | 40 | 11 | 55 | 20 | 100 | 2,50 |
| Did you find portability featureof "touchscreen writing/drawing system"useful? | 2 | 10 | 7 | 35 | 11 | 55 | 20 | 100 | 2,45 |
| Do you think that being able to teach your lesson while sitting is a useful feature? | 1 | 5 | 6 | 30 | 13 | 65 | 20 | 100 | 2,60 |
| Do you think writing on a tuuchscreeen is as easy as writing on a board? | 3 | 15 | 6 | 30 | 11 | 55 | 20 | 100 | 2,40 |
| **Overall Mean:** | 1,8 | 9 | 6,6 | 33 | 11,6 | 58 | 20 | 100 | 2,49 |
| No (1) 1,00 – 1,66 %0  –  %33<br>Partly (2) 1,67 – 2,33 %34  –  %66<br>Yes (3) 2,34 – 3,00 %67  –  %100 | | | | | | | | | |

When Table 6 is examined in terms of the items the followings can be inferred: touchscreen is usable to teach lesson; the features of the system are necessary and used appropriately; being portable increases the usability of the system; and it is as usable as blackboards since it can be used to teach while sitting.

The graphs showing mean values regarding "usability" factor are presented under Figure 13.

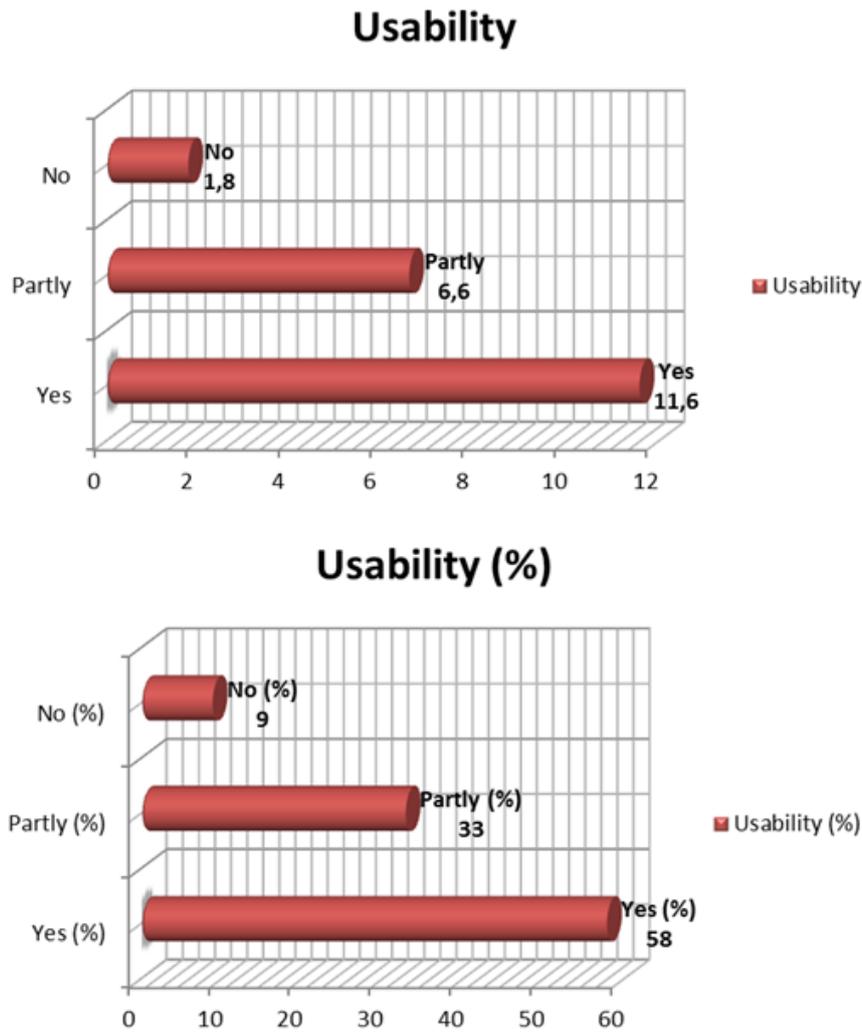

**Figure 13.** Usability survey study – The mean values regarding the findings of "usability" evaluation.

## 6. CONCLUSIONS AND FUTURE WORK

The main objective of the current study is that the touchscreen writing / drawing system designed and developed within the framework of this study will supplement education process. In addition, this system is expected to contribute to "educational technology" field in terms of extending the field and giving it a multi – directional feature. In this respect, the authors try to introduce the system as an efficient and effective one so as to integrate education and technology.

The literature review carried out prior to the study revelaed that there is not adequate number of materials and teaching tools that are likely to contribute to increase the quality of activities used in teaching and to reduce the problems faced by the teachers while teaching. Therefore; the "touchscreen writing/drawing system" designed and developed in the current study is thought to be a solution to the needs mentioned.

In parallel with the study, software and hardware design process of this system was explained without mentioning the complex technical details. In addition, the study was evaluated in general first by asking the teachers, who are the real target users of the system, to use the system and then analyzing the feedback provided by them. Thanks to this evaluation approach, which involves usability process and the survey study following this process, the authors tried to determine to what extent the system was effective and efficient and the objectives were met.

When the observations and the data obtained during evaluation process are considered, it is clear that the touchscreen writing / drawing system is used by teachers in a positive way and therefore is likely to contribute to classroom teaching processes by increasing the effectiveness and efficiency of the lessons for both teachers and students. At this point, the usability process applied revealed that touchscreen could be used in a positive way and is an effective tool to provide solutions to the limitations mentioned before. The survey study carried out after the usability process aimed at evaluating "touchscreen writing/drawing system" with regards to three variables, namely "subservientess, user – friendliness and usability", and the feedback received regarding these factors revealed positive opinions about the sytem. As a result, "touchscreen writing/drawing system" developed within the framework of this study is thought to be effective material – device in meeting the objectives of the study and to contribute the related literature to a great extent.

The positive findings and feedback obtained in this study are quite encouraging for the authors to conduct further studies on the topic. Accordingly, improving software and hardware infrastructure of "touchscreen writing/drawing system" is planned within the scope of these further studies. In this regard, certain studies for improving the system are planned to have a more effective, comprehensive and interactive software in the future versions of the system. Finally, the following actions are also planned in the future: to develop interfaces to enable the interaction with different systems; to use high storage memory units; and to develop touch panels in different sizes.

**ACKNOWLEDGMENT**

Patent application of the designed and developed touchscreen writing / drawing system was processed by the Turkish Patent Institute (Application No: 2013/01298) and supported by The Scientific and Technological Research Council of Turkey (TUBITAK).

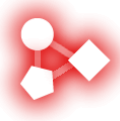

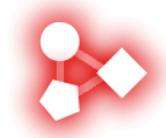